\documentclass[11pt,twoside]{article}

\usepackage{asp2006}
\usepackage{epsf}
\usepackage{psfig}
\usepackage{lscape}

\newcommand{\tel}{ASKAP}
\newcommand{\hi}{H\,\textsc{i}}
\newcommand{\HI}{H\,{\sc i}}

\begin{document}
\title{Science with the Australian Square Kilometre Array Pathfinder (ASKAP)}   %%% Fill in title
\author{Simon Johnston, Ilana J. Feain and Neeraj Gupta}   %%% Fill in author names
\affil{Australia Telescope National Facility, CSIRO, Australia}    %%% Fill in author affiliations

\begin{abstract} %%% Abstract to run on from here.
The future of cm and m-wave astronomy lies with the
Square Kilometre Array (SKA), a telescope under development by a 
consortium of 17 countries that will be 50 times more sensitive than 
any existing radio facility.
Most of the key science for the SKA will be addressed through large-area
imaging of the Universe at frequencies from a few hundred MHz to a few GHz.
The Australian SKA Pathfinder (\tel) is a technology demonstrator aimed
in the mid-frequency range, and achieves instantaneous wide-area imaging
through the development and deployment of phased-array feed systems on
parabolic reflectors.
The large field-of-view makes \tel\ an unprecedented synoptic
telescope that will make substantial advances in SKA key science.
\tel\ will be located at the Murchison Radio 
Observatory in inland Western Australia, one of
the most radio-quiet locations on the Earth and one of
two sites selected by the international community as a potential
location for the SKA.
In this paper, we outline the \tel\ project and summarise its headline science goals as defined by the community at large.

\end{abstract}

%%% MAIN BODY OF TEXT GOES HERE. CONSULT "INSTRUCTIONS FOR AUTHORS USING
%%% LATEX2E MARKUP", SECTIONS 2.3-2.6 FOR HELP WITH EQUATIONS, FIGURES,
%%% AND TABLES.

\section{Introduction}   %%% Top level section head (remove "%" symbol)
The Australian SKA Pathfinder (\tel) is a next generation radio
telescope on the strategic pathway towards the staged development 
of the Square Kilometre Array (SKA). 
In this short conference paper an overview of \tel\  is given and 
a brief description of some indicative science programs. An expanded 
write-up of the  \tel\ science case can be found in the 
refereed papers (Johnston et al. 2007 and Johnston et al. 2008). Details and updates for 
the \tel\ project generally can be found on the 
web at http://www.atnf.csiro.au/projects/askap.

\noindent \tel\ has four main goals:
\begin{itemize}
\item to demonstrate and prototype technologies for the mid-frequency
SKA, including field-of-view enhancement by focal-plane phased-arrays
on new-technology 12-m class parabolic reflectors;
\item to carry out world-class, ground breaking observations directly
relevant to the SKA Key Science Projects;
\item to establish a site for radio astronomy in Western Australia where
observations can be carried out free from the harmful effects of 
radio interference;
\item to establish a user community for the SKA.
\end{itemize}

\section{\tel\ System Parameters}
Table~\ref{intro:params} gives the \tel\ system parameters. The first column gives the relevant parameter with the second column listing
the value of that parameter. 
\begin{table}
\begin{center}
\caption{System parameters for \tel. Note that the field of view is frequency independent.}
\begin{tabular}{lc}
\hline & \vspace{-3mm} \\
\multicolumn{1}{c}{Parameter} & Value\\
\hline & \vspace{-3mm} \\
Number of Dishes & 36 \\
Dish Diameter (m) & 12\\
Total collecting area (m$^2$) & 4072\\
Aperture Efficiency & 0.8  \\
System Temperature (K) & 50 \\
Field-of-view (deg$^2$) & 30 \\
Frequency range (MHz) & 700 $-$ 1800\\
Instantaneous Bandwidth (MHz) & 300 \\
Maximum number of channels & 16384  \\
Configuration maximum baseline (m) & 6000\\
\hline & \vspace{-3mm} \\
\end{tabular}
\label{intro:params}
\end{center}
\end{table}

\vspace*{-10mm}
\subsection{Survey Speeds}
There are two important metrics for telescopes like \tel. The first
is the survey speed; the amount of sky which can be surveyed in a given
time to a given sensitivity. The second is the instantaneous sensitivity.
The first metric can be expressed as a proportionality
$\propto A^2\,T^{-2}\,F$, where $A$ is the total (effective) collecting
area (m$^{2}$), $T$ the system temperature (K) and $F$ the field 
of view (sq deg). Instantaneous sensitivity is proportional to
$A\,T^{-1}$. The survey speed for \tel\ is
1.3$\times$10$^5$~m$^4$K$^{-2}$deg$^2$ and the sensitivity is
65~m$^2$K$^{-1}$. This is a few percent that expected for the final
SKA design.

In Table~\ref{intro:speed} the sensitivities and survey speeds for 
different angular resolutions are given. The first entry gives a 
continuum survey speed where the entire 300~MHz of bandwidth is 
exploited and a desired 1-$\sigma$ sensitivity of 100~$\mu$Jy is 
required. The second entry lists the continuum sensitivity for a one 
hour observation. The third entry gives a spectral line survey 
speed with a spectral resolution of 100kHz and a desired 
sensitivity of 5\,mJy, with the fourth listing the line sensitivity for a 
one hour observation. The fifth line gives the surface brightness 
survey speed for a 1-$\sigma$ limit of 1~K over 5~kHz channel and the 
final entry the surface brightness sensitivity 
reached in a one hour observation. 

\subsection{Array Configuration}
A number of science projects (pulsar surveys, Galactic \hi,  low 
surface brightness mapping) require a highly compact array 
configuration  in order to increase the surface brightness survey speed.
On the other hand science such as continuum and transients require 
long baselines both to overcome the effects of confusion and to 
obtain accurate positions for identification at other wavelengths.
In the middle is the extragalactic \HI\ survey which needs moderate 
resolution to avoid over-resolving the sources. With a total of 36 dishes,
it is difficult to achieve all these requirements simultaneously. The 
initial ASKAP array configuration (Gupta et al. 2008) has 28 antennas arranged 
within a circle of diameter $\sim$2km, with a further 8 antennas 
placed up to $\sim$3km from the configuration centre yielding a maximum 
baseline of $\sim$6km. This configuration takes into account a 
mask of the Murchison Radio Observatory site. This configuration is 
optimised to produce excellent sensitivity and a good point spread 
function at an angular resolution of 30" at 1.4 GHz. The configuration 
also provides high angular speed at an angular resolution of 10" and 
good surface brightness accuracy. This configuration is expected to 
return excellent science outcomes for ASKAP for at least the first 
five years of science operations. Future upgrade paths consisting of 
antenna reconfiguration or more antennas generally will be contingent on 
future resources and will be driven by community desire. 
\begin{table*}
\begin{center}
\caption{Sensitivity and survey speeds for \tel\ for different angular resolutions, assuming a 50\,K system temperature.}
\begin{tabular}{lcccccc}
\hline & \vspace{-3mm} \\
\multicolumn{1}{c}{Parameter} & 10'' & 18'' & 30'' & 90'' & 180'' & unit\\
\hline & \vspace{-3mm} \\
Continuum survey speed (300 MHz, 100~$\mu$Jy) & 220 & 361  & 267 & 54 & 17 & deg$^2$/hr\\
Continuum sensitivity (300 MHz, 1\,hr) & 37 & 29  & 34 & 74 & 132 & $\mu$Jy/beam\\
Line survey speed (100~kHz, 5~mJy) & 184 & 301&223&45&14 & deg$^2$/hr\\
Line sensitivity (100~kHz, 1hr) & 2.1 & 1.6&1.9&4.1&7.3 & mJy/beam\\
Surface brightness survey speed (5~kHz, 1~K) & -- & -- & 1.1&18&94 & deg$^2$/hr\\
Surface brightness sensitivity (5~kHz, 1hr) & -- & -- & 5.2&1.3&0.56 & K\\
\hline & \vspace{-3mm} \\
\end{tabular}
\label{intro:speed}
\end{center}
\end{table*}

\vspace*{-10mm}
\subsection{Location of ASKAP}
The central core of \tel\ will be located at the 
Murchison Radio  Observatory in inland Western Australia, one of 
the most radio-quiet locations on the Earth and one of 
only two sites short-listed by the international community as a 
potential location for the SKA; the second site is located in Southern Africa.
The approximate geographical coordinates of the site are 
longitude 116.5 east and latitude 26.7 south.
%The southern latitude of \tel\ implies that the Galactic Centre 
%will transit overhead and the Magellanic Clouds will be 
%prominent objects of study. At least 30,000 square degrees of 
%sky will be visible to \tel.
The choice of site ensures that \tel\ will be largely free of 
the harmful effects of radio interference currently plaguing the 
present generation of telescopes, especially at frequencies around 
1~GHz and below.
%Being able to obtain a high 
%continuous bandwidth at low frequencies is 
%critical to much of the science planned for \tel.

\subsection{Timeline }
In early 2008, the \tel\ test bed antenna was installed at the site
of the 64-m Parkes radio telescope to allow testing of the focal plane
array and beamforming systems. In late 2008, the antenna contract for the 
36 \tel\ dishes was signed.
The first 6 antennas of \tel\ will be on-site and operational in
Western Australia by late 2010 and constitute the Boolardy Engineering
Test Array (BETA). Over the subsequent two years
the remainder of the antennas will be deployed, with commissioning
of the final system expected to take place in late 2012.

\section{Science with \tel}
Following international science meetings held in Australia in April
2005 and March 2007, seven main science themes have been identified 
for \tel. These are
extragalactic \HI\ science, continuum science, polarization science,
Galactic and Magellanic science, VLBI science, pulsar science and
the radio transient sky. In this paper, we can do no more than 
list bullet points for each science section. An extended version of
the science case has already been published (Johnston et al. 2007), with
larger, and more complete, version of the 
science case for \tel\  published in Experimental Astronomy (Johnston et al. 2008).

The technological innovation of \tel\ and the unique radio quiet location in
Western Australia will enable a powerful synoptic survey instrument that
will  make substantial advances in SKA technologies and on three of 
the SKA key science projects:
the origin and evolution of cosmic magnetism, the evolution of
galaxies and large scale structure, and strong field tests of gravity.
The headline science goals for \tel\ are:
\begin{itemize}
\item The detection of a million galaxies in atomic hydrogen emission
across 80\% of the sky out to a redshift of 0.2 to understand galaxy
formation and gas evolution in the nearby Universe 

\item The detection of synchrotron radiation from 60 million galaxies to
determine the evolution, formation and population of galaxies across
cosmic time and enabling key cosmological tests.
\item The detection of polarized radiation from over 500,000 galaxies,
allowing a grid of rotation measures at $10'$  to explore
the evolution of magnetic fields in galaxies over cosmic time.
\item The understanding of the evolution of the interstellar medium of
our own Galaxy and the processes that drive its chemical and physical
evolution.
\item The characterization of the radio transient sky through
detection and monitoring of transient sources such as gamma ray
bursts, radio supernovae and intra-day variables.
\item The discovery and timing of up to 1000 new radio pulsars to find
exotic objects and to pursue the direct detection of gravitational waves.
\item The high-resolution imaging of intense, energetic phenomena
through improvements in the Australian and global Very Long Baseline networks.
\end{itemize}

\subsection{Expressions of Interest for Survey Science}
Expressions of Interest (EoI) for survey science projects closed on December 15, 2008. A total
of 38 EoIs were received, and information on the EoIs can be obtained from our web pages.
Teams submitting successful EoIs have been invited to submit a full proposal by June 15, 2009.
Anyone wishing to join teams and participate in ASKAP science should contact the ASKAP Project Scientists by sending email to atnf-askap-ps@atnf.csiro.au

\end{document}